\documentclass[twocolumn]{revtex4}%
\usepackage{graphicx}
\usepackage{amsmath}
\usepackage{amsfonts}
\usepackage{amssymb}%
\providecommand{\U}[1]{\protect\rule{.1in}{.1in}}

\newtheorem{proposition}{Proposition}

\newenvironment{proof}[1][Proof]{\noindent\textbf{#1.} }{\ \rule{0.5em}{0.5em}}
\begin{document}
\title{Attainment of the Multiple Quantum Chernoff Bound for Certain Ensembles of
Mixed States}
\author{Michael Nussbaum}
\affiliation{Department of Mathematics, Cornell University, Ithaca NY 14853, USA }

\begin{abstract}
We consider the problem of detecting the true quantum state among $r$ possible
ones, based on measurements performed on $n$ of copies of a finite dimensional
quantum system. It is known that the exponent for the rate of decrease of the
averaged error probability cannot exceed the multiple quantum Chernoff bound
(MQCB) defined as the worst case (smallest) quantum Chernoff distance between
any possible pair of the $r$ states. This error exponent is attainable for $r$
pure states, but for the general case of mixed states only attainability up to
a factor $1/3$ is known. Here we show that the MQCB is attainable for mixed
states if there is a pair which is closer in quantum Chernoff distance than
$1/6$ times the distance between all other pairs.

\end{abstract}
\maketitle







\section{ Introduction and main result}

Consider the problem of discrimination between several quantum hypotheses
$H_{i}:\rho=\rho_{i}$, $i=1,\ldots,r$, $r\geq2$, where $\Sigma=\left\{
\rho_{1},\ldots,\rho_{r}\right\}  $ is a set of $d\times d$ density matrices
identified with a quantum state on $\mathbb{C}^{d}$. A quantum decision rule
with $r$ possible outcomes is given by a POVM (positive operator valued
measure), that is a set of complex self-adjoint positive matrices $d\times d$
matrices $E=\left\{  E_{1},\ldots,E_{r}\right\}  $ satisfying $\sum_{i=1}%
^{r}E_{i}=\mathbf{1}$. We will refer to the $r$-tuple $E$ as a \textsl{quantum
multiple test} or a \textsl{quantum detector.} In the special case where all
$E_{i}$ are projections, the $r$-tuple $E$ is called a PVM (projection valued
measure) or von Neumann measurement. The individual success probability, i.e.
the probability to accept hypothesis $H_{i}$ when $\rho_{i}$ is the true
state, is given by \textrm{tr }$[\rho_{i}E_{i}]$, with corresponding error
probability \textrm{tr} $[\rho_{i}(\mathbf{1}-E_{i})]$. The total (averaged)
sucess and error probabilities are then
\begin{align*}
\text{\textrm{Succ}}(E)  &  :=\frac{1}{r}\sum_{i=1}^{r}\mathrm{tr\ }[\rho
_{i}E_{i}],\\
\text{\textrm{Err}}(E)  &  :=1-\text{\textrm{Succ}}(E)=\frac{1}{r}\sum
_{i=1}^{r}\text{\textrm{tr} }[\rho_{i}(\mathbf{1}-E_{i})].
\end{align*}
For the case of two hypotheses $r=2$, the optimal (Bayes) test for each
$n\in\mathbb{N}$ is known to be the \textsl{Holevo-Helstrom hypothesis test}.
It is given by the PVM $E^{\dagger}=\left\{  E_{1}^{\dagger},E_{2}^{\dagger
}\right\}  $ where
\begin{equation}
E_{1}^{\dagger}=\mathrm{supp}\ (\rho_{1}^{\otimes n}-\rho_{2}^{\otimes n}%
)_{+},\;E_{2}^{\dagger}=\mathbf{1}-E_{1}^{\dagger}, \label{HH-PVM}%
\end{equation}
$\mathrm{supp}\ a$ being the projection onto the space spanned by the columns
of $a$, and $a_{+}$ denotes the positive part of a self-adjoint operator $a$.
The Bayes detector $E^{\dagger}$ for the general case $r\geq2$ has been
described in \cite{Holevo-JMA-73}, \cite{Yuen}; explicit expressions for its
$r$ components are not known in general if $r>2$.

The above describes the basic setup where the finite dimension $d$ is
arbitrary and the hypotheses are equiprobable. We consider the quantum analog
of having $n$ independent identically distributed observations. For this, the
$r$ hypotheses are assumed to be given by the set $\Sigma^{\otimes
n}:=\left\{  \rho_{1}^{\otimes n},\ldots,\rho_{r}^{\otimes n}\right\}  $
$i=1,\ldots,r,$ where $\rho^{\otimes n}$ is the $n$-fold tensor product of
$\rho$ with itself. The detectors $E=\left\{  E_{1},\ldots,E_{r}\right\}  $
now operate on the states $\rho_{i}^{\otimes n}$, but $E_{i}$ need not have
tensor product structure. The corresponding total error probability of a
detector $E$ is now
\[
\mathrm{Err}_{n}(E)=1-\sum_{i=1}^{r}\frac{1}{r}\text{\textrm{tr} }\left[
\rho_{i}^{\otimes n}E_{i}\right]  .
\]
If for a sequence of detectors $E_{(n)}$ the limit $\lim_{n\rightarrow\infty
}-\frac{1}{n}\log\mathrm{Err}_{n}\left(  E_{(n)}\right)  $ exists, we refer to
it as the \textsl{(asymptotic) error exponent. } For two density matrices
$\rho_{1}$ and $\rho_{2}$ the \textsl{quantum Chernoff bound }is defined by
\begin{equation}
\xi_{QCB}(\rho_{1},\rho_{2}):=-\log\inf_{0\leq s\leq1}\mathrm{tr}\text{
}\left[  \rho_{1}^{1-s}\rho_{2}^{s}\right]  . \label{def:q-Chernoff-dist}%
\end{equation}
The basic properties of $\xi_{QCB}(\rho_{1},\rho_{2})$ have been discussed in
\cite{ANSV}. For the binary discrimination problem, it is known that the
Holevo-Helstrom (Bayes) detector $E_{(n)}^{\dagger}$ satisfies
\begin{equation}
\lim_{n\rightarrow\infty}-\frac{1}{n}\log\mathrm{Err}_{n}\left(
E_{(n)}^{\dagger}\right)  =\xi_{QCB}(\rho_{1},\rho_{2}), \label{binary-QCB}%
\end{equation}
thus specifying $\xi_{QCB}(\rho_{1},\rho_{2})$ as the optimal error exponent
(cf. \cite{Audenaert}, \cite{ANSV}, \cite{NSz}), and providing the quantum
analog of the classical Chernoff bound.

For a set $\Sigma=\left\{  \rho_{1},\ldots,\rho_{r}\right\}  $ of density
operators on $\mathbb{C}^{d}$, where $r\geq2$, the \textsl{multiple quantum
Chernoff bound} (MQCB) $\xi_{QCB}\left(  \Sigma\right)  $ was introduced in
\cite{NSz2}: \emph{ }
\begin{equation}
\xi_{QCB}(\Sigma):=\min\{\xi_{QCB}(\rho_{i},\rho_{j}):\ 1\leq i<j\leq r\}.
\label{MQCB-def}%
\end{equation}
\emph{ } If all the states are jointly diagonizable (commuting), then
(\ref{MQCB-def}) reduces to the classical multiple Chernoff bound, as it was
defined in \cite{Salikh-73}, \cite{Salikh-2002} for hypotheses represented by
probability distributions. The following statement summarizes the known facts
about $\xi_{QCB}(\Sigma)$ as an upper bound on the rate exponent, and its
attainability in the general case of mixed states \cite{NSz2}, \cite{mult}%
.\newpage

\begin{proposition}
\label{prop:known-facts} Let $\Sigma=\left\{  \rho_{1},\ldots,\rho
_{r}\right\}  $ be a finite set of hypothetic states on $\mathbb{C}^{d}$.
\newline a) For any sequence $\{E_{(n)}\}$ of quantum detectors relative to
$\Sigma^{\otimes n}$ one has
\begin{equation}
\limsup\limits_{n\rightarrow\infty}-\frac{1}{n}\log\mathrm{Err}_{n}\left(
E_{(n)}\right)  \leq\xi_{QCB}(\Sigma). \label{ineq:NSz2}%
\end{equation}
b) There exists a sequence $\{E_{(n)}^{\ddagger}\}$ of quantum detectors such
that
\begin{equation}
\lim_{n\rightarrow\infty}-\frac{1}{n}\log\mathrm{Err}_{n}\left(
E_{(n)}^{\ddagger}\right)  \geq\frac{1}{3}\xi_{QCB}(\Sigma).
\label{inequ:attain-1/3}%
\end{equation}

\end{proposition}

Note that (\ref{inequ:attain-1/3}) implies the same relation for the Bayes
detector $E_{(n)}^{\dagger}$. In special cases the factor $1/3$ in
(\ref{inequ:attain-1/3}) can be removed, e.g. if $\Sigma$ is a set of
(distinct) pure states \cite{NSz2} or more generally if $\rho_{i}$ are
pairwise linearly independent states \cite{mult}. For pure states in the
context of quantum optics, in a  local operations and classical communication
(LOCC) framework cf. \cite{Nair-et-al}. 

The condition of pairwise linear independence \cite{Eldar}, \cite{mult} does
not allow for full rank density matrices $\rho_{i}$ (faithful states). The
purpose of this note is to show attainability of $\xi_{QCB}(\Sigma)$ under
another special condition, which allows for faithful states. To state it, for
any pair $i<j$, define $\xi_{ij}:=\xi_{QCB}(\rho_{i},\rho_{j})$ and the
expression
\begin{equation}
\bar{\xi}_{ij}(\Sigma):=\min\{\xi_{kl}:\ 1\leq k<l\leq r,\left(  k,l\right)
\neq\left(  i,j\right)  \}.\label{xi-bar-def}%
\end{equation}

\textbf{Theorem. }\emph{Assume there is a pair of states }$\left\{  \rho
_{i},\rho_{j}\right\}  \subset\Sigma$\emph{, }$i<j$\emph{, such that }%
\begin{equation}
\xi_{ij}\leq\frac{1}{6}\bar{\xi}_{ij}(\Sigma). \label{cond-attain-a}%
\end{equation}
\emph{Then there exists a sequence }$\{E_{(n)}\}$\emph{ of quantum detectors
relative to }$\Sigma^{\otimes n}$\emph{ such that }%
\begin{equation}
\lim_{n\rightarrow\infty}-\frac{1}{n}\log\mathrm{Err}_{n}\left(
E_{(n)}\right)  \geq\xi_{QCB}(\Sigma). \label{attain-MQCB}%
\end{equation}

Note that under the condition of the Theorem, one has $\xi_{QCB}\left(
\Sigma\right)  =\xi_{ij}$, thus $\left\{  \rho_{i},\rho_{j}\right\}  $ is a
unique "least favorable pair", that is, the closest pair in Chernoff distance.
The condition says that that there is a pair with distance smaller than $1/6$
the distance between all other pairs.

\section{Proof of the Theorem}

In what follows we assume a general set $\Sigma$ of $r\geq3$ density
operators, not necessarily fulfilling (\ref{cond-attain-a}). We assume w.l.g.
that the pair $\rho_{1},\rho_{2}$ is a least favorable, so that $\xi
_{QCB}(\Sigma)=\xi_{12}=\min_{i<j}\xi_{ij}$. Note that in general the least
favorable pair need not be unique. To ease notation, we will work with sums
(rather than averages) of success and error probabilities $\mathrm{Err}%
_{\mathrm{sm}}(E):=r\mathrm{Err}(E)$ and $\mathrm{Succ}_{\mathrm{sm}%
}(E):=r\mathrm{Succ}(E)$.

\begin{proposition}
There exists a sequence $\{E_{(n)}\}$ of quantum detectors relative to
$\Sigma^{\otimes n}$ such that
\begin{equation}
-\frac{1}{n}\log\mathrm{Err}_{\mathrm{sm}}(E^{(n)})\geq\min\left(  \xi
_{12},\frac{1}{6}\bar{\xi}_{12}(\Sigma)\right)  . \label{claim-propos}%
\end{equation}
Consequently (\ref{attain-MQCB}) holds if (\ref{cond-attain-a}) is fulfilled.
\end{proposition}

For the proof, we initially assume formally $n=1$ and construct a POVM
relative to $\Sigma$ which uses the Holevo-Helstrom PVM for the pair $\rho
_{1},\rho_{2}$ as an ingredient. To this end, let $E_{i}$, $i=3,\ldots,r$ be
an arbitrary collection of positive operators on $\mathbb{C}^{d}$ satisfying
$\sum_{i=3}^{r}E_{i}\leq\mathbf{1}$, $\sum_{i=3}^{r}E_{i}\neq\mathbf{1}$. In
the POVM\ to be constructed, the $E_{i}$, $i\geq3$ will be understood as the
elements corresponding to a decision in favor of $\rho_{i}$. We will
complement this to a full POVM\ for discriminating between all $\rho_{i},$
$1\leq i\leq r$ in the following way.

Let $E^{\dagger}=\left\{  E_{1}^{\dagger},E_{2}^{\dagger}\right\}  $ be the
Holevo-Helstrom PVM given by (\ref{HH-PVM}) for discriminating between
$\rho_{1}$ and $\rho_{2}$. Note the easily verifiable relations%
\begin{equation}
\rho_{1}E_{2}^{\dagger}+\rho_{2}E_{1}^{\dagger}=E_{2}^{\dagger}\rho_{1}%
+E_{1}^{\dagger}\rho_{2}=:\rho_{1}\wedge\rho_{2} \label{HH-relation}%
\end{equation}
defining a self adjoint, but not necessarily positive matrix $\rho_{1}%
\wedge\rho_{2}$ with the property $\mathrm{tr}\left[  \rho_{1}\wedge\rho
_{2}\right]  =\mathrm{Err}_{\mathrm{sm}}\left(  E^{\dagger}\right)  \geq0$.
Let $\tilde{E}_{3}=\sum_{i=3}^{r}E_{i}$, let $Q=\mathbf{1}-\tilde{E}_{3}$ and
define the full POVM $E=\left\{  E_{1},E_{2},E_{i},i=3,\ldots,r\right\}  $ now
by setting
\[
E_{i}:=Q^{1/2}E_{i}^{\dagger}Q^{1/2},\;i=1,2.
\]
Indeed this is a POVM: $E_{i}$, $i=1,2$ are positive and
\begin{align*}
E_{1}+E_{2}  &  =Q^{1/2}\left(  E_{1}^{\dagger}+E_{2}^{\dagger}\right)
Q^{1/2}\\
&  =Q=\mathbf{1}-\tilde{E}_{3}=\mathbf{1}-\sum_{i=3}^{r}E_{i}.
\end{align*}

\textbf{Lemma.} \textit{With the above determination of a POVM\ }$E,$\textit{
we have }%
\begin{eqnarray*}
	\mathrm{Err}_{\mathrm{sm}}(E)
	&\leq& 
	2\mathrm{tr}\left[\rho_{1}\wedge\rho_{2}\right]
	\\
	& &
	\quad
	+2\mathrm{tr}
	\left[
		\left(\rho_{1}+\rho_{2}\right)
		\left(\sum_{i=3}^{r}E_{i}\right)
  	\right]
	\\
	& &
	\qquad
	+\sum_{i=3}^{r}\mathrm{tr}
	\left[
		\rho_{i}\left(\mathbf{1}-E_{i}\right)
	\right].
\end{eqnarray*}

\begin{proof}
Define $F_{i}:=\mathbf{1}-E_{i}^{\dagger}$, $i=1,2$ (thus $F_{1}%
=E_{2}^{\dagger}$) ; then for $i=1,2$
\begin{align*}
\mathrm{tr}\left[  \rho_{i}E_{i}\right]   &  =\mathrm{tr}\left[  \rho
_{i}Q^{1/2}E_{i}^{\dagger}Q^{1/2}\right] \\
&  =\mathrm{tr}\left[  \rho_{i}Q^{1/2}\left(  \mathbf{1}-F_{i}\right)
Q^{1/2}\right] \\
&  =\mathrm{tr}\left[  \rho_{i}Q\right]  -\mathrm{tr}\left[  \rho_{i}%
Q^{1/2}F_{i}Q^{1/2}\right] \\
&  =\mathrm{tr}\left[  \rho_{i}\right]  -\mathrm{tr}\left[  \rho_{i}\tilde
{E}_{3}\right]  -\mathrm{tr}\left[  \rho_{i}Q^{1/2}F_{i}Q^{1/2}\right] \\
&  =1-\mathrm{tr}\left[  \rho_{i}\tilde{E}_{3}\right]  -\mathrm{tr}\left[
\rho_{i}Q^{1/2}F_{i}Q^{1/2}\right]  .
\end{align*}
Hence
\begin{eqnarray*}
	& & \mathrm{Succ}_{\mathrm{sm}}(E)
	\\
	&=& 
	\sum_{i=1,2}\mathrm{tr}\left[\rho_{i} E_{i}\right]  
	+\sum_{i=3}^{r}\mathrm{tr}\left[\rho_{i}E_{i}\right]
	\\
	&=& 
	2
	-\mathrm{tr}\left[
		\left(\rho_{1}+\rho_{2}\right)\tilde{E}_{3}
	\right]
	\\
	& & 
	-\sum_{i=1,2}\mathrm{tr}\left[\rho_{i}Q^{1/2}F_{i}Q^{1/2}\right]
	+\sum_{i=3}^{r}\mathrm{tr}\left[  \rho_{i}E_{i}\right]
	\\
	&=& 
	r-\mathrm{tr}\left[
	\left(\rho_{1}+\rho_{2}\right)\tilde{E}_{3}
	\right]
	\\
	& & 
	-\sum_{i=1,2}\mathrm{tr}\left[\rho_{i}Q^{1/2}F_{i}Q^{1/2}\right]
	-\sum_{i=3}^{r}\mathrm{tr}\left[
		\rho_{i}\left(\mathbf{1}-E_{i}\right)
	\right].
\end{eqnarray*}

This implies
\begin{align*}
\mathrm{Err}_{\mathrm{sm}}(E)  &  =r-\mathrm{Succ}_{\mathrm{sm}}(E)\\
&  =\mathrm{tr}\left[  \left(  \rho_{1}+\rho_{2}\right)  \tilde{E}_{3}\right]
+\sum_{i=1,2}\mathrm{tr}\left[  \rho_{i}Q^{1/2}F_{i}Q^{1/2}\right] \\
&  +\sum_{i=3}^{r}\mathrm{tr}\left[  \rho_{i}\left(  \mathbf{1}_{d}%
-E_{i}\right)  \right]  .
\end{align*}
Let $R:=\mathbf{1}-Q^{1/2}$; since $0\leq Q\leq\mathbf{1}$, we also have
$0\leq R\leq\mathbf{1}$. Hence
\begin{align}
&  \sum_{i=1,2}\mathrm{tr}\left[  \rho_{i}Q^{1/2}F_{i}Q^{1/2}\right]
\nonumber\\
&  =\mathrm{tr}\left[  \rho_{1}F_{1}\right]  +\mathrm{tr}\left[  \rho_{2}%
F_{2}\right] \label{key-1a-plus}\\
&  -2\mathrm{tr}\left[  \rho_{1}RF_{1}\right]  -2\mathrm{tr}\left[  \rho
_{2}RF_{2}\right] \label{key-2a}\\
&  +\mathrm{tr}\left[  \rho_{1}RF_{1}R\right]  +\mathrm{tr}\left[  \rho
_{2}RF_{2}R\right]  . \label{key-4a}%
\end{align}
In view of (\ref{HH-relation}), expression (\ref{key-1a-plus}) equals
$\mathrm{tr}\left[  \rho_{1}\wedge\rho_{2}\right]  $. Regarding (\ref{key-2a}%
), we have
\begin{align*}
&  \mathrm{tr}\left[  \rho_{1}RF_{1}\right]  +\mathrm{tr}\left[  \rho
_{2}RF_{2}\right] \\
&  =\mathrm{tr}\left[  RF_{1}\rho_{1}\right]  +\mathrm{tr}\left[  RF_{2}%
\rho_{2}\right] \\
&  =\mathrm{tr}\left[  \left(  \rho_{1}\wedge\rho_{2}\right)  R\right]
\end{align*}
and thus for the modulus of (\ref{key-2a}),
\begin{align*}
2  &  \left\vert \mathrm{tr}\left[  \left(  \rho_{1}\wedge\rho_{2}\right)
R\right]  \right\vert \\
&  \leq2\sum_{i=1,2}\left\vert \mathrm{tr}\left[  F_{i}\rho_{i}^{1/2}\rho
_{i}^{1/2}R\right]  \right\vert \\
&  \leq2\sum_{i=1,2}\left(  \mathrm{tr}\left[  F_{i}\rho_{i}F_{i}\right]
\mathrm{tr}\left[  R\rho_{i}R\right]  \right)  ^{1/2}%
\end{align*}
by the Cauchy-Schwarz inequality. Using the inequality $2ab\leq a^{2}+b^{2}$
we deduce
\begin{align*}
&  2\left\vert \mathrm{tr}\left[  \left(  \rho_{1}\wedge\rho_{2}\right)
R\right]  \right\vert \\
&  \leq\sum_{i=1,2}\left(  \mathrm{tr}\left[  F_{i}\rho_{i}F_{i}\right]
+\mathrm{tr}\left[  R\rho_{i}R\right]  \right) \\
&  =\sum_{i=1,2}\left(  \mathrm{tr}\left[  F_{i}\rho_{i}\right]
+\mathrm{tr}\left[  \rho_{i}R^{2}\right]  \right) \\
&  =\mathrm{tr}\left[  \rho_{1}\wedge\rho_{2}\right]  +\mathrm{tr}\left[
\left(  \rho_{1}+\rho_{2}\right)  R^{2}\right]
\end{align*}
using the fact that $F_{i}$ are projections. Note that for any $x\in
\lbrack0,1]$ we have $\left(  1-\left(  1-x\right)  ^{1/2}\right)  ^{2}\leq
x$; hence
\begin{equation}
R^{2}=\left(  \mathbf{1}-\left(  \mathbf{1}-\tilde{E}_{3}\right)
^{1/2}\right)  ^{2}\leq\tilde{E}_{3}. \label{R-squared-inequ}%
\end{equation}
Hence the term (\ref{key-2a}) is bounded in absolute value by
\[
2\left\vert \mathrm{tr}\left[  \left(  \rho_{1}\wedge\rho_{2}\right)
R\right]  \right\vert \leq\mathrm{tr}\left[  \rho_{1}\wedge\rho_{2}\right]
+\mathrm{tr}\left[  \left(  \rho_{1}+\rho_{2}\right)  \tilde{E}_{3}\right]  .
\]
For the term (\ref{key-4a}) we obtain
\begin{align*}
&  \sum_{i=1,2}\mathrm{tr}\left[  \rho_{i}RF_{i}R\right] \\
&  =\sum_{i=1,2}\mathrm{tr}\left[  \rho_{i}^{1/2}RF_{i}R\rho_{i}^{1/2}\right]
\\
&  \leq\sum_{i=1,2}\mathrm{tr}\left[  \rho_{i}^{1/2}R^{2}\rho_{i}^{1/2}\right]
\\
&  \leq\mathrm{tr}\left[  \left(  \rho_{1}+\rho_{2}\right)  \tilde{E}%
_{3}\right]
\end{align*}
where in the last inequality (\ref{R-squared-inequ}) has been used again.
Summarizing the upper bounds for (\ref{key-1a-plus})-(\ref{key-4a}) we obtain
the lemma.
\end{proof}

In view of the decomposition of the error probability given by the Lemma, the
strategy is now to choose a good POVM $\left\{  Q,E_{i},i=3,\ldots,r\right\}
$ for decision between $\left(  \rho_{1}+\rho_{2}\right)  /2$ and $\rho_{i},$
$i=3,\ldots,r$. We will proceed to the tensor product case where $\rho_{i}$ is
replaced by $\rho_{i}^{\otimes n}$. Furthermore, set $n=n_{1}+n_{2}$ where
$n_{i}$ will be determined later. Then $\rho_{i}^{\otimes n}=\rho_{i}^{\otimes
n_{1}}\otimes\rho_{i}^{\otimes n_{2}}$ .

For $i=1,2$, let $E^{(n,i)}:=\left\{  E_{i}^{(n,i)},E_{j}^{(n,i)}\text{,
}j=3,\ldots,r\right\}  $ be an arbitrary POVM for decision between the $r-1$
density operators $\left\{  \rho_{i}^{\otimes n},\rho_{j}^{\otimes
n},j=3,\ldots,r\right\}  $. The corresponding sum of error probabilities is,
for $i=1,2.$
\begin{align*}
\mathrm{Err}_{\mathrm{sm}}(E^{(n,i)})  &  =\mathrm{tr}\left[  \rho
_{i}^{\otimes n}\left(  \mathbf{1}-E_{i}^{(n,i)}\right)  \right] \\
&  +\sum_{j=3}^{r}\mathrm{tr}\left[  \rho_{j}^{\otimes n}\left(
\mathbf{1}-E_{j}^{(n,i)}\right)  \right]  \text{. }%
\end{align*}
We now set
\begin{equation}
E_{j}^{(n)}:=E_{j}^{(n_{1},1)}\otimes E_{j}^{(n_{2},2)},\text{ }j=3,\ldots,r;
\label{current-E-j-def}%
\end{equation}
this choice determines $\tilde{E}_{3}=\sum_{i=3}^{r}E_{j}^{(n)}$ and hence the
full POVM. To estimate the error probability $\mathrm{tr}\left[  \frac{1}%
{2}\left(  \rho_{1}^{\otimes n}+\rho_{2}^{\otimes n}\right)  \tilde{E}%
_{3}\right]  $, consider the two terms separately:
\begin{align*}
&  \mathrm{tr}\left[  \rho_{1}^{\otimes n}\tilde{E}_{3}\right] \\
&  =\sum_{j=3}^{r}\mathrm{tr}\left[  \rho_{1}^{\otimes n}\left(  E_{j}%
^{(n_{1},1)}\otimes E_{j}^{(n_{2},2)}\right)  \right] \\
&  =\sum_{j=3}^{r}\mathrm{tr}\left[  \rho_{1}^{\otimes n_{1}}E_{j}^{(n_{1}%
,1)}\otimes\rho_{1}^{\otimes n_{2}}E_{j}^{(n_{2},2)}\right] \\
&  =\sum_{j=3}^{r}\mathrm{tr}\left[  \rho_{1}^{\otimes n_{1}}E_{j}^{(n_{1}%
,1)}\right]  \mathrm{tr}\left[  \rho_{1}^{\otimes n_{2}}E_{j}^{(n_{2}%
,2)}\right] \\
&  \leq\sum_{j=3}^{r}\mathrm{tr}\left[  \rho_{1}^{\otimes n_{1}}E_{j}%
^{(n_{1},1)}\right]  =\mathrm{tr}\left[  \rho_{1}^{\otimes n_{1}}\left(
\mathbf{1-}E_{1}^{(n_{1},1)}\right)  \right] \\
&  \leq\mathrm{Err}_{\mathrm{sm}}(E^{(n_{1},1)})
\end{align*}
and analogously
\[
\mathrm{tr}\left[  \rho_{2}^{\otimes n}\tilde{E}_{3}\right]  \leq
\mathrm{Err}_{\mathrm{sm}}(E^{(n_{2},2)})
\]
Now for $3\leq j\leq r$ consider the term $\mathrm{tr}\left[  \rho
_{j}^{\otimes n}\left(  \mathbf{1}-E_{j}^{(n)}\right)  \right]  $ in the
overall error probability given by the lemma. With our current definition of
$E_{j}^{(n)}$ (\ref{current-E-j-def}), we have%

\begin{eqnarray*}
	\mathbf{1}-E_{j}^{(n)}
	&=&
	\left(\mathbf{1}-E_{j}^{(n_{1},1)}+E_{j}^{(n_{1},1)}\right)
	\\
	& &
	\quad
	\otimes
	\left(\mathbf{1}-E_{j}^{(n_{2},2)}+E_{j}^{(n_{2},2)}\right)
	\\
	& &
	\quad\quad
	-E_{j}^{(n)}
	\\
	&=&
	\left(\mathbf{1}-E_{j}^{(n_{1},1)}\right)
	\otimes
	\left(\mathbf{1}-E_{j}^{(n_{2},2)}\right)
	\\
	& & 
	\quad
	+\left(\mathbf{1}-E_{j}^{(n_{1},1)}\right)
	\otimes 
	E_{j}^{(n_{2},2)}
	\\
	& & 
	\quad\quad
	+E_{j}^{(n_{1},1)}
	\otimes
	\left(\mathbf{1}-E_{j}^{(n_{2},2)}\right).
\end{eqnarray*}
Consequently
\begin{eqnarray}
	& &
	\mathrm{tr}
	\left[
		\rho_{j}^{\otimes n}\left(\mathbf{1}-E_{j}^{(n)}\right)
	\right]
	\nonumber\\
	&=&
	\mathrm{tr}
	\left[
		\rho_{j}^{\otimes n_{1}}
		\left(\mathbf{1}-E_{j}^{(n_{1},1)}\right)
	\right]  
	\mathrm{tr}
	\left[
		\rho_{j}^{\otimes n_{2}}
		\left(\mathbf{1}-E_{j}^{(n_{2},2)}\right)
	\right]
	\nonumber\\
	& & 
	\quad
	+\mathrm{tr}
	\left[
		\rho_{j}^{\otimes n_{1}}
		\left(\mathbf{1}-E_{j}^{(n_{1},1)}\right)
	\right]
	\mathrm{tr}
	\left[
		\rho_{j}^{\otimes n_{2}}E_{j}^{(n_{2},2)}
	\right] 
	\nonumber\\
	& &
	\quad\quad
	+\mathrm{tr}
	\left[
		\rho_{j}^{\otimes n_{1}}E_{j}^{(n_{1},1)}
	\right]
	\mathrm{tr}
	\left[
		\rho_{j}^{\otimes n_{2}}
		\left(\mathbf{1}-E_{j}^{(n_{2},2)}\right)
	\right]
	\nonumber\\
	&\leq&
	\mathrm{tr}
	\left[
		\rho_{j}^{\otimes n_{1}}
		\left(\mathbf{1}-E_{j}^{(n_{1},1)}\right)
	\right]
	\nonumber\\
	& &
	\quad
	+\mathrm{tr}
	\left[
		\rho_{j}^{\otimes n_{1}}
		\left(\mathbf{1}-E_{j}^{(n_{1},1)}\right)
	\right]
	\nonumber\\
	& & 
	\quad\quad
	+\mathrm{tr}
	\left[
		\rho_{j}^{\otimes n_{2}}
		\left(\mathbf{1}-E_{j}^{(n_{2},2)}\right)
	\right].
	\nonumber
\end{eqnarray}
Hence the sum of error terms is
\begin{align*}
&  \sum_{j=3}^{r}\mathrm{tr}\left[  \rho_{j}^{\otimes n}\left(  \mathbf{1}%
-E_{j}^{(n)}\right)  \right] \\
&  \leq2\sum_{i=1,2}\sum_{j=3}^{r}\mathrm{tr}\left[  \rho_{j}^{\otimes n_{i}%
}\left(  \mathbf{1}-E_{j}^{(n_{i},i)}\right)  \right] \\
&  \leq2\mathrm{Err}_{\mathrm{sm}}(E^{(n_{1},1)})+2\mathrm{Err}_{\mathrm{sm}%
}(E^{(n_{2},2)}).
\end{align*}
Finally we obtain for our overall POVM $E^{(n)}$, according to the Lemma,
\begin{align}
\mathrm{Err}_{\mathrm{sm}}(E^{(n)})  &  \leq2\mathrm{tr}\left[  \rho
_{1}^{\otimes n}\wedge\rho_{2}^{\otimes n}\right] \nonumber\\
&  +4\sum_{i=1,2}\mathrm{Err}_{\mathrm{sm}}(E^{(n_{i},i)}).
\label{overall-err}%
\end{align}
To evaluate this bound, we now have the choice of $n_{1}$, $n_{2}$ and the two
POVM\ $E^{(n_{1},1)}$, $E^{(n_{2},2)}$. We set
\[
n_{1}=\left[  nw_{1}\right]  \text{, }n_{2}=n-n_{1}\text{ where }w_{1}%
+w_{2}=1.
\]
Let us make a crude choice $w_{1}=w_{2}=1/2$. For the POVM $E^{(n_{i},i)},$
which decides between $\rho_{i}^{\otimes n}$, $\rho_{3}^{\otimes n}%
,\ldots,\rho_{r}^{\otimes n}$, we choose the method that attains $1/3$ of the
Chernoff bound for this set of states, that is we choose the detector
$E_{(n_{i})}^{\ddagger}$ from (\ref{inequ:attain-1/3}), for $i=1,2$. Defining
sets of index pairs
\[
J_{i}=\left\{  (k,l)\in\left\{  i,3,\ldots,r\right\}  ^{\times2},k<l\right\}
,i=1,2
\]
we obtain for $i=1,2$%
\begin{align*}
&  -\frac{1}{n_{i}}\log\mathrm{Err}_{\mathrm{sm}}(E^{(n_{i},i)})\\
&  \geq\frac{1}{3}\min\left\{  \xi_{kl}:(k,l)\in J_{i}\right\}  .
\end{align*}
Now note that with $\bar{\xi}_{12}(\Sigma)$ from (\ref{xi-bar-def}) we have
\[
\min\left\{  \xi_{kl}:(k,l)\in J_{1}\cup J_{2}\right\}  =\bar{\xi}_{12}%
(\Sigma)\text{.}%
\]
Thus, taking into account $n_{i}=n/2\left(  1+o(1)\right)  $, $i=1,2,$ we
obtain
\begin{align*}
&  -\frac{1}{n}\log\left(  \mathrm{Err}_{\mathrm{sm}}(E^{(n_{1},1)}%
)+\mathrm{Err}_{\mathrm{sm}}(E^{(n_{2},2)})\right) \\
&  \geq\frac{1}{6}\bar{\xi}_{12}(\Sigma).
\end{align*}
Thus from (\ref{overall-err}) and the binary quantum Chernoff bound
(\ref{binary-QCB})
\[
-\frac{1}{n}\log\mathrm{tr}\left[  \rho_{1}^{\otimes n}\wedge\rho_{2}^{\otimes
n}\right]  \geq\xi_{12}%
\]
we obtain the claim (\ref{claim-propos}).

\section{Conclusions}

Refined results of this type can be obtained if we optimize the sample size
weights $w_{1},w_{2}$ and /or choose the detectors \ $E^{(n_{i},i)}$, $i=1,2$
from the blocking algorithm ("test between pairs" method) applied in
\cite{exp-decay}. Indeed it has been shown in \cite{exp-decay} that for every
$\varepsilon>0$, there are ensembles $\Sigma$ of mixed states and detectors
$E_{(n)}$ such that%
\[
\lim_{n\rightarrow\infty}-\frac{1}{n}\log\mathrm{Err}_{n}\left(
E_{(n)}\right)  \geq(1-\varepsilon)\xi_{QCB}(\Sigma),
\]
improving the general bound (\ref{inequ:attain-1/3}). Furthermore, the method
applied here, based on the risk decomposition given in the Lemma, may be
applied recursively. This shows that there is a multitude of special
configurations of the set $\Sigma$ of general (possibly mixed) states where
the MQCB is attained, lending further support to the conjecture that it is
attainable in general.


\begin{thebibliography}{99}                                                                                               %


\bibitem {Audenaert}Audenaert, K.M.R., Casamiglia, J., Munoz-Tapia, R., Bagan,
E., Masanes, Ll., Acin, A., and Verstraete, F., \textit{Phys. Rev.
Lett.} \textbf{98}, 160501 (2007)

\bibitem {ANSV}Audenaert, K.M.R., Nussbaum, M., Szko\l a, A., and Verstraete,
F., \textit{Comm. Math. Phys.} \textbf{279} (1), 251-283 (2008)

\bibitem {Eldar}Eldar, Y. ,  \textit{Phys. Review A} \textbf{68}, 052303 (2003)

\bibitem {Holevo-JMA-73}Holevo, A.S., \emph{J. Multivar. Anal.} \textbf{3}
(4), 337-394 (1973)

\bibitem {Nair-et-al}Nair, R., Guha, S. and Tan, S.,  arXiv:1212.2048 (2013)

\bibitem {NSz}Nussbaum, M. and Szko\l a, A., \emph{Ann. Statist.}
\textbf{37} (2), 1040-1057 (2009)

\bibitem {NSz2}Nussbaum, M. and Szko\l a, A., in \emph{Theory of
Quantum Computation, Communication and Cryptography.} 5th Conference, TQC
2010, Leeds, UK. Revised Selected Papers. Lecture Notes in Computer Science,
Vol 6519, van Dam, W., Kendon, V. M., and Severini, S. (Eds.), 1-8,
Springer (2011)

\bibitem {exp-decay}Nussbaum, M. and Szko\l a, A., \emph{J. Math.
Phys.} \textbf{51}, 072203 (2010)

\bibitem {mult}Nussbaum, M. and Szko\l a, A. , \emph{Ann. Statist.} \textbf{39} (6), 3211-3233 (2011)


\bibitem {Salikh-73}Salikhov, N. P. \emph{Dokl. Akad. Nauk SSSR} \textbf{209},
54-57 (Russian, 1973)

\bibitem {Salikh-2002}Salikhov, N. P., \emph{Theory Probab. Appl.}
\textbf{47}  (2), 286-298 (2003)

\bibitem {Yuen}Yuen, H.P., Kennedy, R.S., and Lax, M., \emph{IEEE Trans.
Inform. Theory} \textbf{IT-21} (2), 125-134 (1975)
\end{thebibliography}
\end{document}